# Theoretical kinetic study of thermal unimolecular decomposition of cyclic alkyl radicals

B. Sirjean,[1] P.A. Glaude,[1] M.F. Ruiz-Lopez,[2] R. Fournet[1,*]

1) Département de Chimie Physique des Réactions, Nancy Université - CNRS, 1, rue Grandville, BP 20451, 54001 Nancy Cedex, France

2) Equipe de Chimie et Biochimie Théoriques, SRSMC, Nancy Université - CNRS, Boulevard des Aiguillettes, BP 239, 54506 Vandoeuvre-lès-Nancy, France

e-mail:  Rene.Fournet@ensic.inpl-nancy.fr

**Title running head:** Ring Opening of Cycloalkyl Radicals

**Corresponding author footnotes**: Tel: 33 3 83 17 50 57, Fax: 33 3 83 37 81 20

ABSTRACT: While many studies have been reported on the reactions of aliphatic hydrocarbons, the chemistry of cyclic hydrocarbons has not been explored extensively. In the present work, a theoretical study of the gas-phase unimolecular decomposition of cyclic alkyl radicals was performed by means of quantum chemical calculations at the CBS-QB3 level of theory. Energy barrier and high–pressure limit rate constants were calculated systematically. Thermochemical data were obtained from isodesmic reactions and the contribution of hindered rotors was taken into account. Classical transition state theory was used to calculate rate constants. Tunneling effect was accounted for in the case of C-H bond breaking. Three-parameter Arrhenius expressions were derived in the temperature range of 500 to 2000

K at atmospheric pressure and the C-C and C-H bond breaking were studied for cyclic alkyl radicals with a ring size ranging from 3 to 7 carbon atoms, with or without a lateral alkyl chain. For the ring opening reactions, the results clearly show an increase of the activation energy as the π bond is being formed in the ring (*endo* ring opening) in contrast to the cases in which π bond is formed on the side chain (*exo* ring opening). These results are supported by the analysis of the electronic charge density that have been performed with the theory of Atom in Molecules (AIM). For all cycloalkyl radicals considered, C-H bond breaking exhibits larger activation energies than C-C bond breaking, except for cyclopentyl for which the ring opening and H loss reactions are competitive over the range of temperature studied. Theoretical results compare rather well with experimental data available in the literature. Evans-Polanyi correlations for C-C and C-H β-scissions in alkyl and cycloalkyl free radicals were derived. The results highlight two different types of behavior depending on the strain energy in the reactant.

MANUSCRIPT TEXT

1. **Introduction**

In recent years, many chemical kinetic studies have investigated the oxidation of straight or branched-chain alkanes. In comparison, little attention has been paid to cyclic alkanes and the chemistry involved during their oxidation.[1] However, cycloalkanes and alkylcycloalkanes (in particular $C_5$ and $C_6$) are usually present in conventional fuels (up to 3% in gasoline and 35% in diesel fuel)[2]. Under oxidative conditions, the reaction sequence starts from H-abstraction by $O_2$ and/or from unimolecular dissociation. The latter leads to diradical species which are specific to the chemistry of cycloalkanes.[3] H-abstractions from the parent cycloalkane lead to cycloalkyl radicals. At low-temperatures (between 650 and 850 K) the cyclic radical reacts with $O_2$ to produce a peroxycycloalkyl radical which isomerizes to form hydroperoxycycloalkyl radical. Several papers have studied these important low-temperature pathways[4-7]. At high temperatures, cycloalkyl radicals decompose mainly through β-scission reactions. These processes are well known for straight and branched alkyl radicals, but the rate constants are not well known for cycloalkyl radicals because of complications due to ring strain. In a recent study, Orme *et al.*[8] examined the pyrolysis and oxidation of methylcyclohexane (MCH) over the temperature range of 1200 to 2100 K. They proposed a detailed chemical kinetic mechanism to simulate shock-tube and flow-reactor experiments. High-temperature reactions of Orme *et al.* were considered in a low-temperature mechanism, proposed by Pitz *al.*, to model the auto-ignition of MCH in a rapid compression machine. In these studies, the rate constants for the ring opening reactions of cycloalkyl radicals were estimated from the reverse ring closure process with rates equal to those reported by Matheu *et al.*[9] In fact, Matheu *et al.* had developed rules for ring closures or openings starting from model reactions listed by Newcomb[10] with rate parameters from literature and calculated using quantum chemical methods for cyclobutyl *endo* ring opening, 1-penten-5-yl *endo* ring closure and 1-hexen-6-yl *endo* ring closure. Even though theoretical and experimental studies have been carried out on the modeling of cyclohexane oxidation, no systematic study has been reported for other cycloalkanes and, in particular, alkylcycloalkanes that are commonly present in practical fuels. In general, the kinetic parameters of the ring opening of cycloalkyl radicals and their reverse, the internal addition of radical centers to a double bond, are still poorly known. The influence of the ring strain energy of a cyclic radical on the activation energy is poorly understood. Moreover, alkyl side-chains tend to complicate the kinetics further.

The ring opening of small cyclic alkyl radicals has been investigated theoretically by several authors. For example, the ring opening reaction of cyclopropyl was studied at various ab initio and DFT levels of theory[11,12]. It represents a system small enough to be studied at a reasonably high level of theory. The reaction of cyclopropylcarbinyl radical to allyl radical was investigated.[13,14] The energy barriers for the ring opening of the $C_3$ ring in these two systems are dramatically different: 21.9 kcal/mol for the cyclopropyl radical[12] *vs* 7.1 kcal/mol for the cyclopropylcarbinyl radical.[14] This large effect of a methyl group can be related to the "*endo*" and "*exo*" reaction types proposed by Newcomb.[10] **Scheme 1** illustrates these two different ring opening/closing reactions. In "*exo*" ring opening, the radical center is located on the side chain while in "*endo*" ring opening, the radical center is on the ring itself.

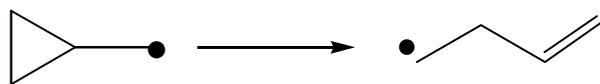

cyclopropylcarbinyl *exo* ring opening

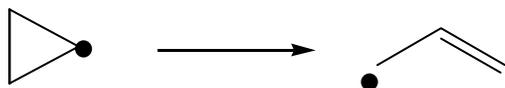

cyclopropyl *endo* ring opening

**Scheme 1**: « *Exo* » and « *Endo* » ring opening reactions for cyclopropylcarbinyl radical and cyclopropyl radical

In order to get a deeper insight on ring opening reactions, we studied C-C and C-H β-scissions for cycloalkyl radicals from $C_3$ to $C_7$ by quantum chemistry calculations. The results were compared to those obtained for straight (or unstrained) free radicals. The differences found are discussed in terms of *endo* and *exo* β-scissions. Moreover, to link the nature of the bonds created or broken in the transition state with those involved in the reactant and product, AIM[15] analyses were carried out. The effect of alkyl substitution on ring opening was also investigated systematically: (a) by varying the size of the ring for a given alkyl side-chain and (b) by varying the size of the alkyl side-chain for a given ring size. In the same way, C-H bond breaking was examined accordingly with an emphasis on the branching ratio of C-C and C-H bond cleavage. Finally, an Evans-Polanyi correlation is proposed for C-C and C-H bond scissions of cyclic and straight-chain alkyl free radicals.

For all species considered in this study, thermochemical data were estimated and the enthalpy of formation $\Delta_f H°_{298}$ was compared with experimental values when available or with estimates of group additivity methods when data are unavailable. Kinetic parameters were determined for all reactions involving $C_4$, $C_5$ and $C_6$ rings. The results are compared with available values in the literature.

## 2. Computational method

Calculations were performed with Gaussian 03 Rev. C.02.[16] The composite method CBS-QB3[17] was applied for all stationary geometries and transition states involved in the reaction schemes. Vibrational frequencies calculated at the B3LYP/cbsb7[18,19] level of theory confirm that all transition states (TS) have exactly one imaginary frequency. The methodology used to obtain thermochemical and kinetic data was described elsewhere.[3] Thermochemical data were derived from CBS-QB3 energy and frequencies. Internal rotors were treated with the *hinderedRotor* option of Gaussian03.[20] It must be stressed that the constrained torsions of the cyclic structure have been treated as harmonic oscillators (including ring floppy motions) and the free alkyl groups as hindered rotations. Enthalpies of formation ($\Delta_f H°$) were obtained using isodesmic reactions. The isodesmic reactions considered as well as the enthalpies of formation of the reference species used can be found in the supporting information.

Spin contamination was only observed for transition states at the CBS-QB3 level of theory (0.8 $<s^2>$ 1.2). It must be noted that, in the CBS-QB3 method, an empirical correction for spin contamination is performed for the energy calculation[17].

Rate constants for each elementary reaction were calculated using classical Transition State Theory (TST).[21] Tunnelling effect was taken into account for C-H bond breaking using the transmission coefficient of Wigner.[22] The enthalpies of activation involved in TST theory were calculated by taking into account the enthalpies of reaction calculated with isodesmic reactions in an elementary unimolecular reaction, reactant (R) → products (P) such as :

$$\Delta H^{\neq}(R \to P) = \left( \Delta H^{\neq}_{1 \, (CBS\text{-}QB3)} + \Delta H^{\neq}_{-1(CBS\text{-}QB3)} + \Delta H_{r \, (isodesmic)} \right)/2 \qquad (1)$$

and

$$\Delta H^{\neq}(P \to R) = \left( \Delta H^{\neq}_{1 \, (CBS\text{-}QB3)} + \Delta H^{\neq}_{-1(CBS\text{-}QB3)} - \Delta H_{r \, (isodesmic)} \right)/2 \qquad (2)$$

where $\Delta H^{\neq}_{1 \, (\mathbf{CBS\text{-}QB3})}$ and $\Delta H^{\neq}_{-1(\mathbf{CBS\text{-}QB3})}$ are, respectively, the enthalpy of activation for the forward and back reactions calculated at a temperature T(K). $\Delta H_{r(isodesmic)}$ corresponds to the enthalpy of reaction calculated at T(K) using NASA polynomial obtained from calculated isodesmic enthalpies of formation and entropy at 298 K, and heat capacities of reactants and products presented in Table 1.

In the text, we often assimilated activation energy with activation enthalpy calculated from TST. This last quantity is expressed as:

$$\Delta^{\neq} H°(T) = \Delta^{\neq} U°(T) + \Delta^{\neq} nRT \quad (3a)$$

and $$E_{exp} = \Delta^{\neq} U°(T) + RT \quad (3b)$$

where $\Delta^{\neq} U°$ takes into account, electronic energy at 0 K, ZPE, and the thermal corrections due to translational, vibrational and rotational contributions to the internal thermal energy. $E_{exp}$ corresponds to the empirical activation energy (Arrhenius activation energy). In the case of unimolecular reactions involved here ($\Delta^{\neq} n = 0$), $\Delta^{\neq} H°$ differs from classical Arrhenius activation energy from the quantity RT.

The kinetic parameters were obtained by fitting the rate constant values obtained from TST at several temperatures between 500 and 2000 K with:

$$k_{\infty} = A\,T^n\,exp\,(-E/RT) \quad (4)$$

where A, n, E are the parameters of the modified Arrhenius equation and $k_{\infty}$ is the high–pressure limit rate constant.

The AIM2000 program[23] was used to carry out the AIM analysis of the electronic charge density.

## 3. Thermochemical data

Thermochemical data ($\Delta_f H°$, $S°$, $C_p°$) for all the species involved in this study are collected in **Table 1**.

As mentioned previously, the enthalpies of formation, $\Delta_f H°$ (column 1), were obtained from isodesmic reactions. The theoretical enthalpies of formation of molecular species are in good agreement with experimental values.[24] The mean absolute deviation from experiment (MAD) given for CBS-QB3 calculations[17] (G2 set) is approximately equal to 0.9 kcal.mol$^{-1}$. By comparison with the experimental enthalpies of formation, found in the literature for stable molecules (Table1), we observe that the differences with CBS-QB3 calculations are always within this uncertainty.

For free radicals, our theoretical values are within a few kcal.mol$^{-1}$ to group additivity results[25] of Thergas.[26] Of course, group additivity does not properly take into account for the ring strain energy (RE) of the radical species for which experimental values are unavailable.

**Table 1**: Ideal gas phase thermodynamics properties for species considered in this study and computed at the CBS-QB3 level. $\Delta H°_{f,298K}$ in kcal.mol$^{-1}$ and $S°_{298K}$ and $C°_p(T)$ in cal.mol$^{-1}$.K$^{-1}$. The last column gives the experimental value of $\Delta H°_{f,298K}$ from NIST[27] or estimated by Thergas[26] (*in italic*).

| Species | $\Delta H°_{f,298}$ | $S°_{298}$ | $C°_p(T)$ | | | | | | | $\Delta H°_{f,298}$ |
| --- | --- | --- | --- | --- | --- | --- | --- | --- | --- | --- |
| | | | 300 K | 400 K | 500 K | 600 K | 800 K | 1000 K | 1500 K | |
|  | 13.0 | 56.8 | 13.3 | 18.0 | 22.2 | 25.7 | 31.0 | 34.9 | 41.0 | 12.7±0.14[28] |
|  | 6.5 | 63.2 | 17.0 | 23.5 | 29.4 | 34.3 | 41.9 | 47.5 | 55.9 | 6.4 |
|  | -17.9 | 70.2 | 21.0 | 29.2 | 36.7 | 43.0 | 52.8 | 59.9 | 70.7 | 18.3±0.19[29] |
|  | -29.4 | 71.5 | 25.4 | 35.3 | 44.3 | 52.0 | 63.9 | 72.5 | 85.7 | -29.8[30] |
|  | -2.2 | 73.5 | 23.1 | 31.0 | 37.9 | 440 | 53.0 | 59.7 | 70.2 | *0.4* |
|  | -25.7 | 81.6 | 27.2 | 36.7 | 45.2 | 52.5 | 63.9 | 72.2 | 85.0 | -25.5[31] |
|  | -30.5 | 90.0 | 32.5 | 41.5 | 53.5 | 61.9 | 75.0 | 84.6 | 99.2 | -30.4±0.25[32] |
|  | 2.6 | 75.4 | 24.9 | 33.7 | 41.6 | 48.2 | 58.2 | 65.4 | 76.4 | 2.4[33] |
|  | 67.8 | 60.9 | 12.4 | 15.9 | 18.9 | 21.4 | 25.0 | 27.7 | 31.8 | *66.0* |
|  | 38.8 | 62.7 | 15.5 | 21.0 | 25.9 | 29.9 | 35.9 | 40.2 | 46.7 | 37.5±0.4[34] |
|  | 8.3 | 69.7 | 19.4 | 26.7 | 33.1 | 38.5 | 46.7 | 52.6 | 61.5 | 8.5[33] |
|  | -0.1 | 72.8 | 23.9 | 32.7 | 40.7 | 47.4 | 57.7 | 65.1 | 76.3 | -1.0±0.23[35] |
| 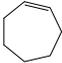 | -1.2 | 79.9 | 28.9 | 39.2 | 48.6 | 56.6 | 68.9 | 77.8 | 91.2 | *-2.1* |
| 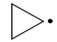 | 69.5 | 61.7 | 13.5 | 17.7 | 21.3 | 24.2 | 28.6 | 31.7 | 36.7 | *66.9* |
| 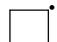 | 54.0 | 68.8 | 17.9 | 23.7 | 28.9 | 33.2 | 39.7 | 44.5 | 51.7 | *50.7* |
| 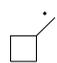 | 46.7 | 78.8 | 22.8 | 30.2 | 36.6 | 42.0 | 50.3 | 56.3 | 65.6 | *48.5* |
| 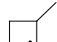 | 46.6 | 76.9 | 23.9 | 31.1 | 37.4 | 42.7 | 50.8 | 56.7 | 65.9 | *44.8* |
| 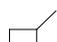 | 46.3 | 77.0 | 23.8 | 31.2 | 37.5 | 42.8 | 50.9 | 56.8 | 66.0 | *44.8* |
| 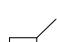 | 44.1 | 78.8 | 22.4 | 29.4 | 35.9 | 41.4 | 50.0 | 56.2 | 65.7 | *44.8* |
| 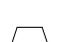 | 25.7 | 71.7 | 21.6 | 29.2 | 36.1 | 41.8 | 50.5 | 56.9 | 66.5 | *23.9* |
| 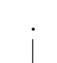 | 23.7 | 85.5 | 26.9 | 35.9 | 43.9 | 50.7 | 61.1 | 68.7 | 80.4 | *22.5* |
| 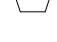 | 16.9 | 81.3 | 26.2 | 35.0 | 43.1 | 50.1 | 60.8 | 68.6 | 80.5 | *16.8* |
| 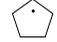 | 18.8 | 79.2 | 27.7 | 36.8 | 44.7 | 51.4 | 61.7 | 69.2 | 80.7 | *16.8* |
| 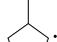 | 18.5 | 79.0 | 27.8 | 36.8 | 44.7 | 51.4 | 61.7 | 69.2 | 80.7 | *16.8* |
| 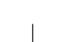 | 12.0 | 90.2 | 31.6 | 42.0 | 51.5 | 59.5 | 71.8 | 80.9 | 94.6 | *11.9* |
| 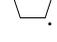 | 17.1 | 77.8 | 26.1 | 35.3 | 43.6 | 50.7 | 61.5 | 69.4 | 81.3 | *18.0* |
| 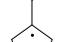 | 15.3 | 84.5 | 31.0 | 42.3 | 52.4 | 60.8 | 73.7 | 82.9 | 96.7 | *11.8* |
| 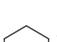 | 39.1 | 63.1 | 14.9 | 18.8 | 22.0 | 24.7 | 28.75 | 31.8 | 36.6 | *40.9±0.7[36]* |
| 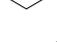 | 49.6 | 75.7 | 21.1 | 26.4 | 31.0 | 34.7 | 40.4 | 44.8 | 51.5 | *48.7* |
| 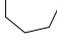 | 44.3 | 86.1 | 24.8 | 31.2 | 37.0 | 41.8 | 49.4 | 55.0 | 63.9 | *43.7* |
| 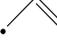 | 43.7 | 85.0 | 25.3 | 31.9 | 37.7 | 42.5 | 49.8 | 55.3 | 64.0 | *41.3* |

| Species | $\Delta H^\circ_{f,298}$ | $S^\circ_{298}$ | $C^\circ_p(T)$ | | | | | | | $\Delta H^\circ_{f,298}$ |
|---|---|---|---|---|---|---|---|---|---|---|
| | | | 300 K | 400 K | 500 K | 600 K | 800 K | 1000 K | 1500K | |
| 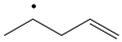 | 41.4 | 85.8 | 24.2 | 30.2 | 35.9 | 40.8 | 48.7 | 54.6 | 63.7 | *41.5* |
| 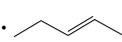 | 41.8 | 84.5 | 25.9 | 31.6 | 37.0 | 41.7 | 49.2 | 54.9 | 63.8 | *41.1* |
| 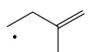 | 41.3 | 85.0 | 26.3 | 32.3 | 37.7 | 42.2 | 49.6 | 55.1 | 63.9 | *40.0* |
| 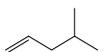 | 37.2 | 93.6 | 31.6 | 39.5 | 46.3 | 52.0 | 61.0 | 67.7 | 78.3 | *36.4* |
| 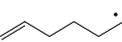 | 39.1 | 92.8 | 29.4 | 37.0 | 44.1 | 50.1 | 59.6 | 66.7 | 77.8 | *38.7* |
| 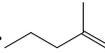 | 36.0 | 91.4 | 31.7 | 39.5 | 46.6 | 52.7 | 62.3 | 69.4 | 80.6 | *35.4* |
| 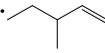 | 37.4 | 92.2 | 31.6 | 39.4 | 46.1 | 51.8 | 63.1 | 67.5 | 78.2 | *37.6* |
| 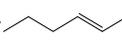 | 36.9 | 91.6 | 30.1 | 37.4 | 44.2 | 50.1 | 59.5 | 66.6 | 77.8 | *37.0* |
| 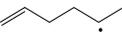 | 36.2 | 91.5 | 29.3 | 36.6 | 43.5 | 49.6 | 59.3 | 66.5 | 77.8 | *36.5* |
| 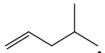 | 37.2 | 93.6 | 31.6 | 39.5 | 46.3 | 52.0 | 61. | 67.7 | 78.3 | *36.4* |
| 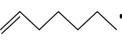 | 33.9 | 101.8 | 35.6 | 44.6 | 52.9 | 59.9 | 70.9 | 79.2 | 92.1 | *33.8* |

Hence, for branched cycloalkyl radicals, the same enthalpy of formation is obtained for isomers with a radical center located on the ring, in contrast to theoretical calculations. This result can be explained by the same bond dissociation energy used in Thergas for these last free radicals. Notice also that the lowest $\Delta H^\circ_{f,298K}$ is obtained for isomers with a radical center located on the tertiary carbon atom. This result is consistent with the variation of bond dissociation energy (BDE): BDE ($C_{primary}$-H) > BDE($C_{secondarry}$-H) > BDE($C_{tertiary}$-H).

### β-scission of non-branched cycloalkyl radicals.

### 4.1 Mechanism.

The ring opening reactions and the C-H bond β-scissions are presented in **Scheme 2**, from cyclopropyl to cycloheptyl.

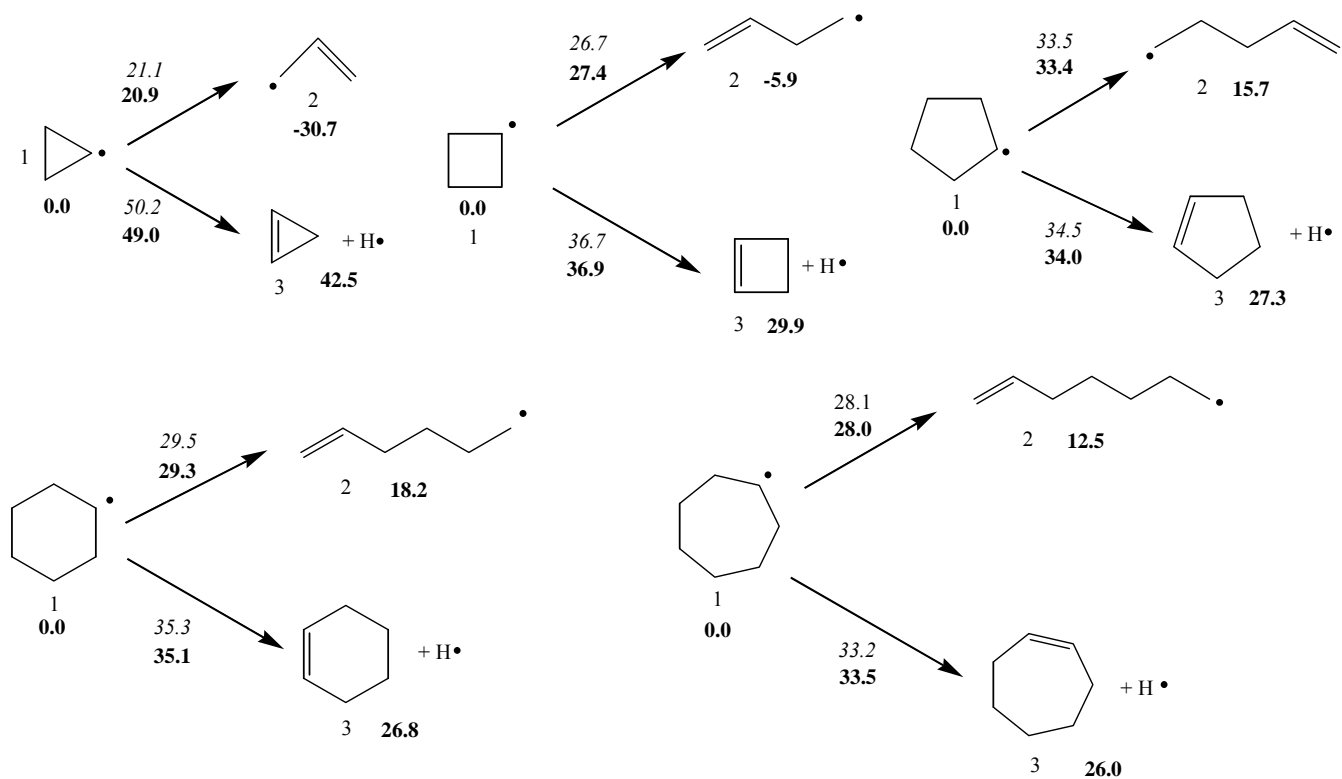

**Scheme 2**: β-scissions of $C_{3-7}$ cycloalkyl radicals, calculated at the CBS-QB3 level of theory. Gibbs free energies (in bold) and activation energies (in italic) are reported in kcal.mol$^{-1}$, at standard state, in relation to the reference cycloalkyl radical.

**Table 2** summarizes the activation enthalpies obtained for the ring opening (and closure) of cycloalkyl radicals (Scheme 2) and the corresponding β-scission of straight-chain 1-alkyl radicals (equation 5) at the CBS-QB3 level of theory.

$$\cdot C_nH_{2n+1} \rightarrow C_2H_4 + \cdot C_{(n-2)}H_{2(n-2)+1} \text{ with } 3 \leq n \leq 7, \tag{5}$$

**Table 2**: Activation enthalpies and enthalpies of reaction ($\Delta_r H^\circ_{298K}$) of the ring opening of cyclic alkyl free radicals described in Scheme 2, and activation enthalpies for the β-scissions of the corresponding linear alkyl free radicals obtained at the CBS-QB3 level of theory (in kcal.mol$^{-1}$ at 298 K).

| Reaction of cyclic radical | $\Delta H^{\circ\neq}_{298K}$ ring opening | $\Delta H^{\circ\neq}_{298K}$ ring closure | $\Delta_r H^\circ_{298K}$ | $\Delta H^{\circ\neq}_{298K}$ β-scission of unstrained n-alkyl (see text) |
|---|---|---|---|---|
| $c$-C$_3$H$_5$ → C$_3$H$_5$ | **21.1** | 51.4 | - 30.3 | **28.9** |
| $c$-C$_4$H$_7$ → C$_4$H$_7$ | **26.7** | 31.4 | - 4.7 | **27.8** |
| $c$-C$_5$H$_9$ → C$_5$H$_9$ | **33.5** | 14.7 | 18.8 | **28.2** |
| $c$-C$_6$H$_{11}$ → C$_6$H$_{11}$ | **29.5** | 7.4 | 22.1 | **28.0** |
| $c$-C$_7$H$_{13}$ → C$_7$H$_{13}$ | **28.1** | 11.0 | 18.0 | **28.0** |

In general, CBS-QB3 activation energies are in good agreement with literature values. The activation energy of cyclopropyl ring opening is close to both the theoretical values of Olivella et al.[12] (21.9 kcal.mol$^{-1}$), Arnold and Carpenter[11] (21.5 kcal.mol$^{-1}$) and Greig and Thynne[37] (22.06 kcal.mol$^{-1}$) but also to the experimental one given by Kerr and al.[38] (19.1 kcal.mol$^{-1}$) at P=0.7 bar and T ranging from 411 to 446 K. $\Delta H^{\circ\neq}_{298K}$ for the ring opening of cyclobutyl is close to that calculated by Matheu et al.[9] at the CBS-Q level of theory (25.9 kcal.mol$^{-1}$). For cyclopentyl and cyclohexyl radicals, Matheu et al.[9] reported the activation energies for the reverse reactions. If we compare their activation energies with our values at 298 K, a good agreement is obtained (15.8 kcal.mol$^{-1}$ *vs* 14.7 kcal.mol$^{-1}$ for $c$-C$_5$H$_9$ and 6.4 kcal.mol$^{-1}$ *vs* 7.4 kcal.mol$^{-1}$ for $c$-C$_6$H$_{11}$).

As mentioned by Stein and Rabinovitch[39], two opposite effects may be considered to explain ring opening energetics. The first one is the ring strain energy being released in the TS leading to reduced activation energy ($E_a$) as compared to unstrained radicals. The second one is due to local orientation strain (steric effect) which tends to increase $E_a$. Since the ring strain energy varies from one cycloalkyl to another, as shown in Table 2, activation energies involved in ring opening differ greatly from one cycloalkyl radical to another. The lowest value is 21.1 kcal.mol$^{-1}$ for the C-C β-scission of cyclopropyl while an activation energy of 33.5 kcal.mol$^{-1}$ is reached for cyclopentyl. On the other hand, the activation energy for C-C bond β-scissions in straight-chain alkyl radicals is found to be independent of the number of carbon atoms involved in the chain, as expected.

The difference between cyclic and non cyclic radicals is clearly associated with the release of ring strain energy (RE) in the former case. In the transition structure (TS) of cyclic radicals, RE is partially released, leading to a lower activation energy.

Some RE values for cycloalkanes and cycloalkenes are available in the literature[40] but no value has been reported for the corresponding radicals. RE values for the molecules cannot be used for the radicals owing to geometry changes. In a previous paper, we showed that CBS-QB3 calculated RE for cycloalkanes are in excellent agreement with experimental ones[3]. Here, RE for the radicals were calculated with a similar approach. RE is obtained by calculating the difference between the enthalpy of formation of the cyclic radical and the sum of contributions to enthalpy of the different groups constituting the cyclic radical and deduced from unstrained structures. For example, in the case of cyclopentyl, RE is obtained from the following calculation:

$$RE = \Delta_f H°_{298K}(cyclopentyl) - 4 \times \Delta_f H°_{298K}(CH_2\ group) - \Delta_f H°_{298K}(\overset{\bullet}{C}H\ group) \qquad (6)$$

The contribution to enthalpy of the CH$_2$ group has been obtained, in a previous study[3] while the contribution of a CH(·) group has been obtained by the difference of the enthalpies of formation ($\Delta_f H°$) between n-hexane and pent-3-yl radical. $\Delta_f H°$ have been obtained from CBS-QB3 calculations and isodesmic reactions. **Table 3** presents the results for the cycloalkyl radicals together with those reported by Cohen for cycloalkanes and cycloalkenes.[40]

**Table 3:** Ring Strain Energies (RE) of cycloalkanes and cycloalkenes (from Cohen[40]) and cylcloalkyl radicals (calculated here at the CBS-QB3 level) for species containing from 3 to 7 carbon atoms (in kcal.mol$^{-1}$).

| Number of carbon atoms in the cycle | RE Cyclanes | RE Cycloalkyl radicals | RE Cycloalkenes |
|---|---|---|---|
| 3 | 27.7 | **38.1** | 53.6 |
| 4 | 26.8 | **24.4** | 29.8 |
| 5 | 7.1 | **4.1** | 5.9 |
| 6 | 0.7 | **1.8** | 0.5 |
| 7 | 6.8 | **4.4** | 5.4 |

The results show that RE of cyclanes, cyloalkyl radicals and cycloalkenes are similar for $C_6$ and $C_7$ species, while substantial variations are observed for smaller species, especially for $C_3$. It is worth noting that RE of cyclopentyl is lower than that of cyclopentane or cyclopentene, which is consistent with the peculiar low reactivity of cyclopentane observed during its pyrolysis and oxidation[41]. The results of Table 3 also indicate that it is generally inaccurate to estimate the activation energy for the β-scission of a cyclic alkyl radical, even as a first approximation, by considering that RE is totally released in the transition state. For example, this assumption yields an activation enthalpy of 27.8 – 24.4 = 3.4 kcal.mol$^{-1}$ for c-$C_4H_7$, which is much smaller than the theoretical value (26.7 kcal.mol$^{-1}$).

The TS involved in the ring opening of cyclopentyl radical is presented in **Figure 1**. All the other TSs in Scheme 2 exhibit similar characteristics.

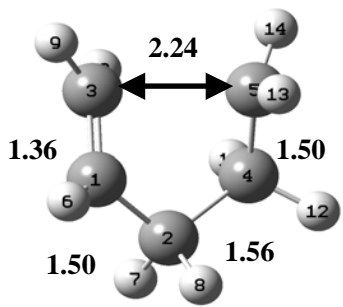

**Figure 1**: Structure of the transition state involved in the β-scission of the C-C bond of cyclopentyl radical at the CBS-QB3 level of calculation. Bond lengths are given in Angstrom.

The TSs are generally tight. In Figure 1, the breaking C-C bond length is equal to 2.24 Å, only 1.4 times the initial bond length in the cyclopentyl radical. Consequently, a part of the RE remains in the TS and does not contribute to decrease the activation energy. On the other hand, the steric inhibition due to the formation of a π bond in the cyclic TS produces an increase in $E_a$. To form a π bond, the $CH_2$ group must rotate in the ring to bring the atoms in same plan (here, atoms 1, 3, 6, 9, 10). This effect is particularly important in the case of cyclopentyl. Its activation energy is 5.3 kcal.mol$^{-1}$ above that of the n-pentyl radical. Thus the steric inhibition due to the nascent π bond strongly affects the activation energy of the ring opening of cyclopentyl radical, to an extent that is not compensated by the partial released of the RE.

### 4.2 AIM analysis

The reaction coordinate can be viewed as a combination of the C-C σ bond breaking and the formation of a π bond. Roughly, one can consider that a decrease in RE is accompanied by σ bond breaking and that the steric inhibition becomes significant as the π bond forms. To characterize this competition, we used AIM to analyse the electronic charge density[15]. This theory is based on a topological analysis of the electronic density ρ. Different critical points can be identified in a molecular structure. Bond Critical Point (BCP) are localized on bond paths that connect two atoms. When bond paths connect atoms in a cycle, a Ring Critical Point (RCP) may be defined in the AIM theory. The value of the electronic density ρ at BCP is characteristic of the bond type. Thus, electronic densities of ρ=0.239 and ρ=0.343 at BCPs characterize, respectively, a single bond and a double bond. **Figure 2** presents the molecular graphs of the different species involved in the ring opening of cyclopentyl radical.

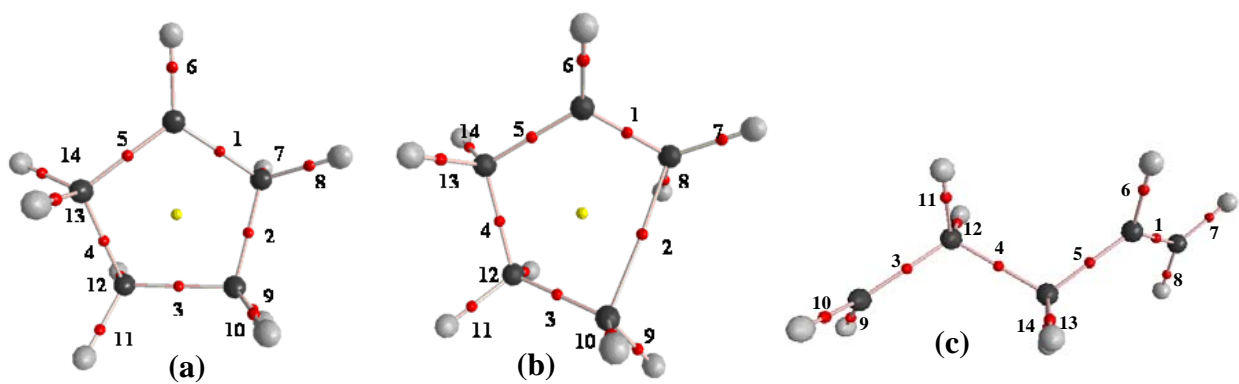

**Figure 2**: Molecular graphs obtained by AIM analysis[23] for the cyclopentyl radical (a), transition state involved in the ring opening of cyclopentyl (b) and 1-pentenyl (c). The bond paths appear in white, BCPs correspond to little dots and the RCP is characterized by the "central" dot.

The density values corresponding to C-C bonds are given in **Table 4** for each structure. In cyclopentyl (Figure 2a), BCP #3 corresponds to a σ bond with a classical value ρ = 0.238. The bonds located in the β position of the radical center (BCP #2 and #4) have a lower density, showing that the bond is weakened; on the contrary, bonds located in the α position of the radical center (BCP #1 and #5) have a density slightly higher than a classical σ bond. In the transition state (Figure 2b), the bond characterized by BCP #2 is clearly being broken.

**Table 4:** Density values, ρ at bond critical points (BCPs) for structures in Figure 2.

| Cyclopentyl (a) | | TS cyclopentyl (b) | | 1-pentenyl (c) | |
|---|---|---|---|---|---|
| BCP number | ρ | BCP number | ρ | BCP number | ρ |

| 1 | 0.255 | 1 | 0.321 | 1 | 0.343 |
| 2 | 0.234 | 2 | 0.053 | 2 | -     |
| 3 | 0.238 | 3 | 0.251 | 3 | 0.258 |
| 4 | 0.234 | 4 | 0.227 | 4 | 0.224 |
| 5 | 0.255 | 5 | 0.251 | 5 | 0.254 |

The small electronic density at this BCP ($\rho = 0.053$) indicates that a bonding character remains in the TS. BCP #1 increases from 0.255 to 0.321, indicating the bond in the TS is close to that of a standard double bond. Thus, even though a part of RE is released at the TS, a substantial structural deformation accompanying $\pi$ bond formation contributes to an increase in the activation energy (compared to $\beta$-scissions of straight-chain alkyl radicals). In a similar way, one can now explain why for high strained cycloalkyl radicals, such as cyclopropyl, for which the RE realeased is expected to dominate over the steric inhibition, the activation energy is smaller than that observed in the $\beta$-scission of the corresponding straight-chain alkyl radical (21.1 kcal.mol$^{-1}$ for cyclopropyl versus 28.9 kcal.mol$^{-1}$ for n-propyl).

We also performed a systemic AIM analysis for all cycloalkyl ring openings discussed in section 4 and for the $\beta$-scissions of the corresponding straight-chain 1-alkyl (equation 5). For each reaction, we were interested in the variations of the electronic density at BCPs located on the $\alpha$ and $\beta$ position of the radical center. We introduce an empirical parameter, $\lambda$, defined by:

$$\lambda = \frac{\rho(BCP(\alpha))}{\rho(BCP(\beta))} \tag{7}$$

For all the cycloalkyl and linear 1-alkyl radicals containing from 4 to 7 carbon atoms, $\lambda$ has been calculated for the reactants and the TSs (**Table 5**).

**Table 5**: Values of λ calculated from equation (7) for the ring openings and the β-scissions of $C_4$ to $C_7$ species.

| Species | Reactant | | | TS | | |
|---|---|---|---|---|---|---|
| | $\rho(BCP(\alpha))$ | $\rho(BCP(\beta))$ | λ | $\rho(BCP(\alpha))$ | $\rho(BCP(\beta))$ | λ |
| 1-alkyl radicals ($C_4$ to $C_7$) | 0.257 | 0.229 | **1.1** | 0.326 | 0.045 | **7.2** |
| Cyclobutyl | 0.253 | 0.227 | **1.1** | 0.309 | 0.067 | **4.6** |
| Cyclopentyl | 0.255 | 0.234 | **1.1** | 0.321 | 0.053 | **6.1** |
| Cyclohexyl | 0.257 | 0.235 | **1.1** | 0.327 | 0.045 | **7.3** |
| Cycloheptyl | 0.254 | 0.232 | **1.1** | 0.326 | 0.046 | **7.1** |

For 1-alkyl radicals, a perfect regularity of λ is observed for all the reactants (λ =1.1) and the TSs (λ = 7.2). For cycloalkyl free radicals, one notes that λ is still equal to 1.1, independent of the ring size. These results indicate that the strengthening of the α bond and the destabilisation of the β bond are similar for these radicals. For cyclic TSs, a bond character is always observed in the transition states ($\rho(BCP(\beta)) > 0$), but λ varies greatly from $C_4$ to $C_7$ species. In fact, the λ values for cyclobutyl and cyclopentyl radicals are smaller than those of their straight-chain analogs, which suggests that the TSs are closed to the reactants and that a large part of the ring strain energy is retained. On the other hand, the λ value for cyclohexyl and cycloheptyl are similar to those of their straight-chain analogs, what may be explained by the similar reactivity exhibited by these radicals towards C-C bond cleavage, as mentioned above.

### 4.3 Kinetic data

Kinetic parameters were calculated according to equation (3) for $C_4$, $C_5$ and $C_6$ rings, which are the most relevant ones in pyrolysis and combustion. The results are presented in **Table 6**.

**Table 6**: Rate parameters for C-C and C-H bond breaking obtained at the CBS-QB3 level of theory and the classical transition state theory. $500 \leq T\ (K) \leq 2000$.

| Reaction | A (s$^{-1}$) | n | E (kcal.mol$^{-1}$) |
|---|---|---|---|
| c-C$_4$H$_7$ → C$_4$H$_7$ | 4.36×10$^{11}$ | 0.539 | 26.84 |
| c-C$_4$H$_7$ → c-C$_4$H$_6$ + H | 2.00×10$^{11}$ | 1.001 | 37.14 |
| c-C$_5$H$_9$ → C$_5$H$_9$ | 4.79×10$^{12}$ | 0.570 | 34.43 |
| c-C$_5$H$_9$ → c-C$_5$H$_8$ + H | 2.95×10$^{12}$ | 0.847 | 35.42 |
| c-C$_6$H$_{11}$ → C$_6$H$_{11}$ | 2.75×10$^{12}$ | 0.624 | 30.81 |
| c-C$_6$H$_{11}$ → c-C$_6$H$_{10}$ + H | 8.91×10$^{11}$ | 0.834 | 36.34 |

Only few rate constant values are available in the literature for the ring opening of cycloalkyl radicals. In the following, we compare our values with those found in the NIST kinetics database[27]. **Figure 3** compares the calculated values of the cyclobutyl ring opening with those proposed by Matheu et *al.*[9] As shown, the rate constants of the two studies agree within a factor 1.2 to 1.7 in the range of 500K to 2000K.

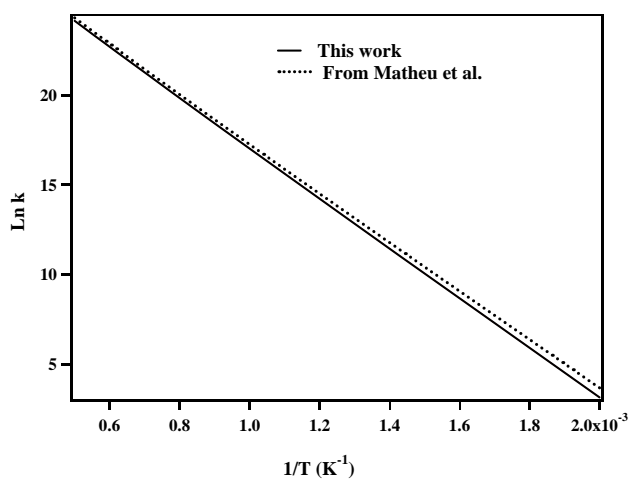

**Figure 3**: Comparison between of the rate constant for the reaction $c$-C$_4$H$_7$ → C$_4$H$_7$ calculated at the CBS-QB3 level (this work) and that of Matheu et al.[9], at the CBS-Q level of theory.

Figure 4a presents a comparison between the rate constant calculated for the ring opening of cyclopentyl radical and the experimental values reported by Hanford-Styring and Walker[42] and Gierzak et al.[43].

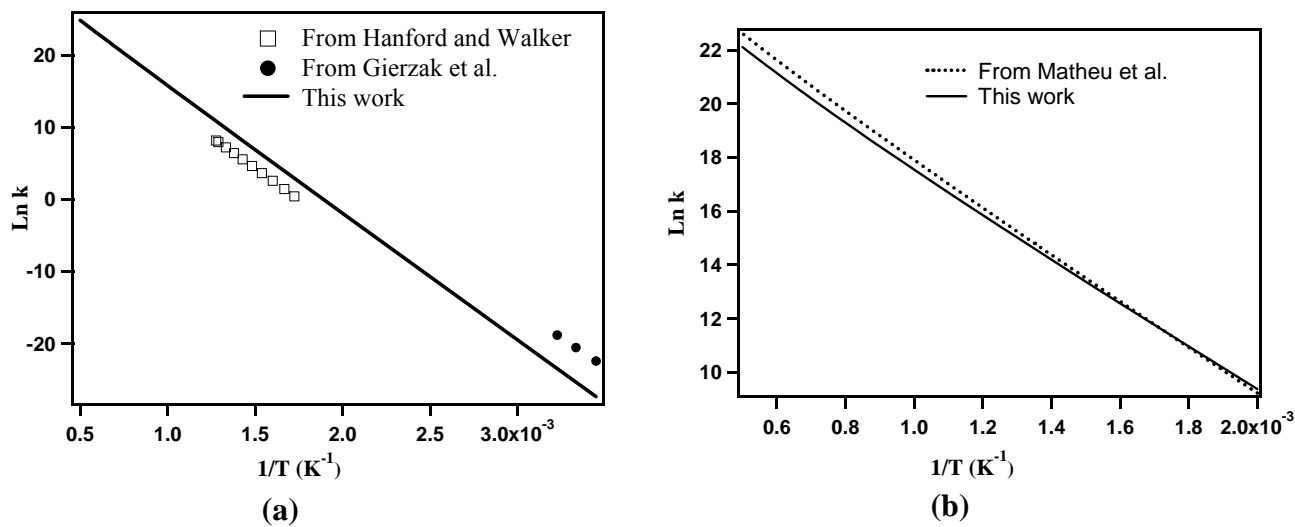

**(a)**                              **(b)**

**Figure 4** : Comparison between rate constants of this work and (a) experimental data for the reaction $c\text{-}C_5H_9 \rightarrow C_5H_9$ and (b) those of Matheu et al.[9] for the ring closure reaction $C_5H_9 \rightarrow c\text{-}C_5H_9$.

For ring opening, our rate constant is with a factor 12 higher than the values of Handford-Styring and Walker (derived from a complex mechanism). In a recent study on the mechanism of decomposition of pentenyl radicals and pressure effects, Tsang[44] proposed a high-pressure rate expression for the cyclic ring opening of cyclopentyl based on the values proposed by Handford-Styring and Walker. The same discrepancy is observed with values proposed by Tsang, ten times lower than ours. Even larger discrepancy is observed with the work of Gierczak et al.[43] since there is a factor greater than 100 between our values and those derived by these authors from chemical activation experiments and RRKM calculations employing experimental data of their work as well as literature data. In the low temperature range considered in the study of Gierczak et al. (around 400 K), their estimated activation energy (approximately 2 kcal.mol$^{-1}$ greater than ours) and their pre-exponential factor twice higher than that calculated from CBS-QB3, yield to the large discrepancy observed. Matheu et al.[9] have studied the reverse reaction, i.e., the intra addition $C_5H_9 \rightarrow$ $c$-$C_5H_9$. **Figure 4b** presents a comparison of our rate constant for the latter reaction (**Table 7**) with that calculated by Matheu et *al*. As shown in Figure 4b, an excellent agreement is obtained, with a factor of 1 to 1.7 between the two rate constants.

**Table 7**: Rate parameters for intra addition $C_5H_9 \rightarrow$ $c$-$C_5H_9$ and $C_6H_{11} \rightarrow$ $c$-$C_6H_{11}$, obtained at the CBS-QB3 level of theory and the classical transition state theory. $500 \leq T~(K) \leq 2000$.

| Reaction | A (s$^{-1}$) | n | E (kcal.mol$^{-1}$) |
|---|---|---|---|
| $C_5H_9 \rightarrow$ $c$-$C_5H_9$ | $3.73 \times 10^6$ | 1.391 | 14.31 |
| $C_6H_{11} \rightarrow$ $c$-$C_6H_{11}$ | $5.31 \times 10^4$ | 1.921 | 6.67 |

**Figure 5** presents the rate constants for the ring closure reaction leading to cyclohexyl radical, $C_6H_{11} \rightarrow$ $c$-$C_6H_{11}$. For comparison, rate constant calculated by Matheu et *al*.[9] and that estimated by Handford-Styring and Walker[42] are also presented.

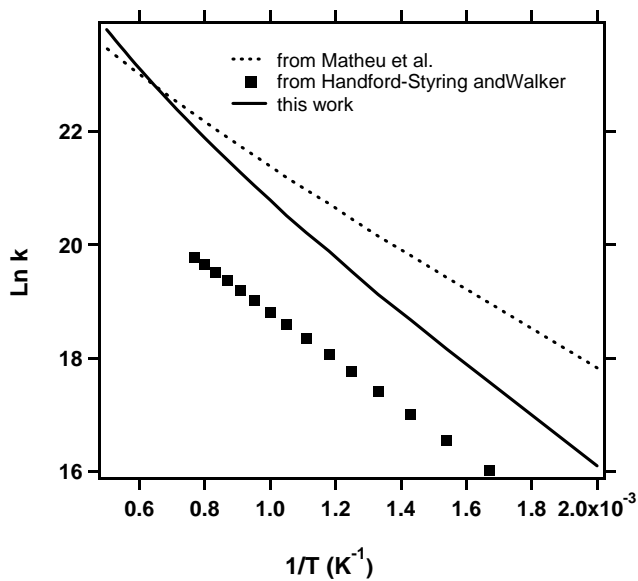

**Figure 5**: Comparison between calculated rate constant in this work and values in the literature for the reaction $C_6H_{11} \rightarrow$ $c$-$C_6H_{11}$.

In contrast to C$_4$ and C$_5$ rate constants, a poor agreement is observed between our values and those calculated by Matheu et *al*. In particular, at 500 K, their value is about 6 times higher than ours and the slopes of the two curves are somewhat different. However, the deviation observed is not surprising because these authors have used a lower computational level (B3LYP/ccp-VTZ) for the intra addition of 1-hexenyl than that used in the cases of cyclobutyl and cyclopentyl radicals (CBS-Q) and consequently their rate constant seems to be overestimated in the low temperature range. The rate values obtained in our calculations are higher than those obtained by Handford-Styring and Walker (derived from a complex mechanism) by a factor ranging from 5 to 10. It is worth noting that a large discrepancy exists between these two rate constants available in the NIST kinetic data base[27] and that the kinetic parameters proposed by Matheu et *al*. have been used by several authors[4,8,45] in the modeling of cyclohexane combustion.

The ratio between C-C and C-H bond cleavage for species considered in Scheme 2 constitutes another interesting point. Due to the difference between energy barrier involved in the C-C and C-H bond scission of cyclopropyl and cyclobutyl and to a lesser extent of cyclohexyl and cycloheptyl, the ratio is always shifted towards the ring opening, even at high temperature. However, for cyclopentyl radical, the difference in activation energy observed for C-C and C-H rupture is relatively small and the two channels have competitive rate constants, as shown in **Figure 6**.

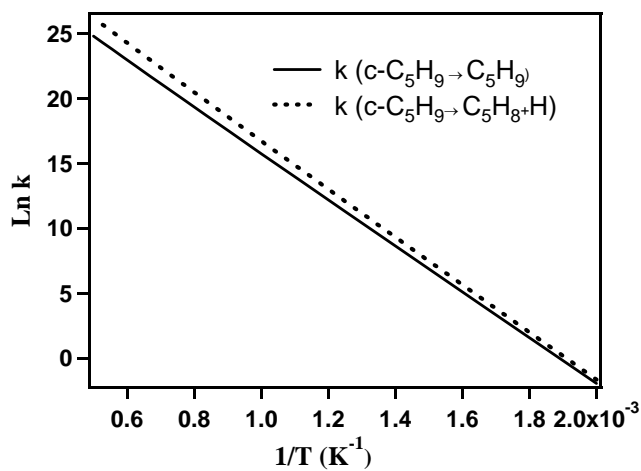

Figure 6: Comparison between the rate constants of C-C and C-H β-scission for cyclopentyl radical (See, Scheme 2 and Table 5).

It is seen that above 500 K, C-H β-scission is faster than C-C scission. A consequence of this result is that $C_5$ cyclic structures exhibit a different reactivity during combustion processes compared to other cyclic species like cyclohexane[41].

5. **Branched cycloalkyl radicals decomposition**

   5.1 **Methylcycloalkyls decomposition mechanism**

As mentioned in the introduction, the *exo* ring opening of cyclopropylcarbinyl has a lower activation energy than the *endo* ring opening of cyclopropyl radical, respectively, 7.06 kcal.mol$^{-1}$ [14], and 21.9 kcal.mol$^{-1}$ [12]. In scheme 1, only *endo* ring opening can be involved during the β-scission of the C-C bonds and it could be interesting to add an alkyl group on the cycle to allow *exo* ring opening. The effect of methyl substitution on the ring is studied in detail here for methylcylobutyl and methylcyclopentyl. The radical center can be located on the ring itself or on the methyl side chain. Radicals formed by H-abstraction on methylcyclobutane and methylcyclopentane are presented, respectively, in **Schemes 3** and **4**. The radicals formed by C-C bond breaking and the activation energies for the corresponding reactions are also displayed.

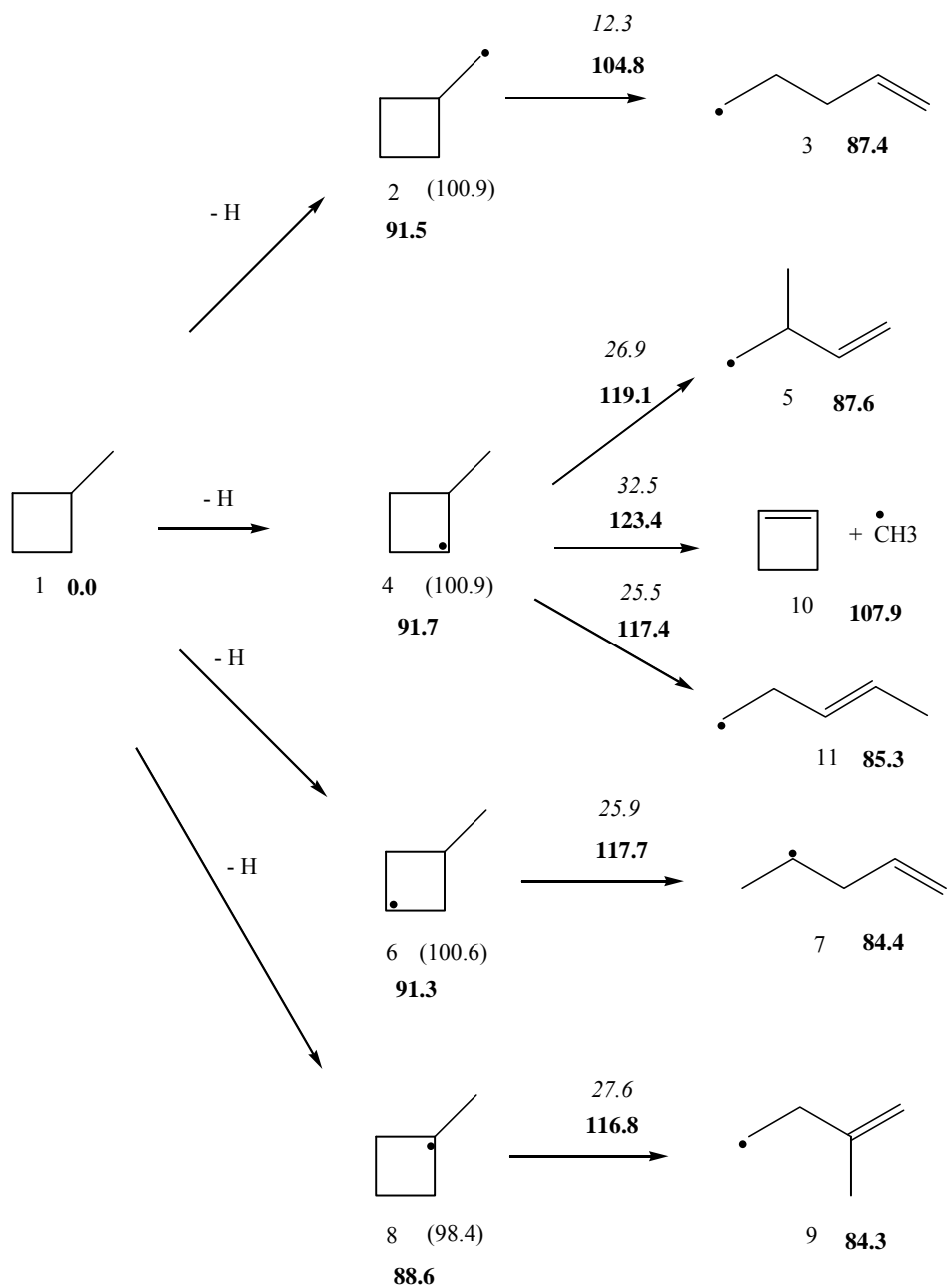

**Scheme 3**: β-scissions for the different isomers of methylcyclobutyl radicals. Gibbs free energies (in bold), activation enthalpies (in italic) and bond dissociation energies (in parentheses) are reported at 298 K, in kcal.mol$^{-1}$ and are relative to the reference cycloalkane.

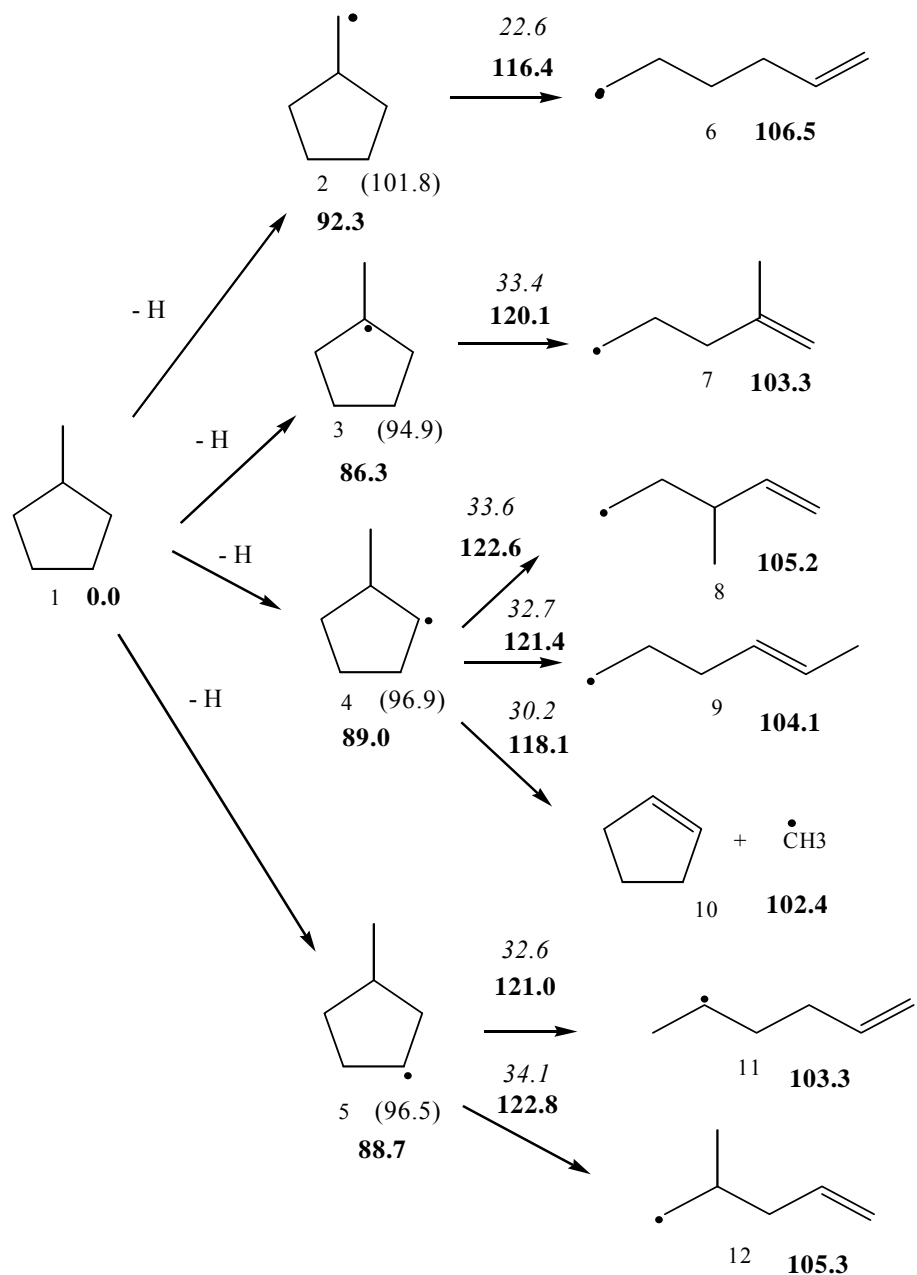

**Scheme 4**: β-scissions for the different isomers of methylcyclopentyl radicals. Gibbs free energies (in bold), activation enthalpies (in italic) and bond dissociation energies (in parentheses) are reported at 298 K, in kcal.mol$^{-1}$ and are relative to the reference cycloalkane.

The enthalpies of reaction displayed in Schemes 3 and 4 correspond to the C-H bond dissociation energies (BDE). For methylcyclobutane, BDEs found for the secondary carbon atoms (radicals 4 and 6 of Scheme 3) are close to the value proposed by Tumanov and Denisov[46] and equal to 100.0 kcal.mol$^{-1}$.

For the primary carbon atom (radical 2 of Scheme 3), our value is close to the BDE recommended by Luo[47] for primary C-H bonds in *n*-butane (100.7 kcal.mol$^{-1}$). For tertiary carbon atom (radical 8 of Scheme 3) our value is 3 kcal.mol$^{-1}$ higher than the BDE of a tertiary C-H bond in *iso*-butane (95.7 kcal.mol$^{-1}$)[47] which could be used, by analogy, to model this BDE in methylcyclobutane. In the case of methylcyclopentane (Scheme 4), our calculations reproduce the classical stability of C-H bonds in primary, secondary and tertiary carbon atoms, observed in linear alkyl radicals[47]. Thus, for the primary carbon atom (radical 2, Scheme 4), our BDE is close to that given for a primary C-H bond in *n*-pentane (100.2 kcal.mol$^{-1}$)[47] while the BDE values for the two secondary carbon atoms (radicals 4 and 5 of Scheme 4) agree with those for cyclopentane (95.6 kcal.mol$^{-1}$)[47]. For the tertiary C-H bond (radical 3 of Scheme 4), the BDE is close to that observed for iso-butane[47].

In Schemes 3 and 4, three types of C-C bond dissociation are considered for methylcyclobutyl and methylcylopentyl radicals:

- The β-scission of the methyl group, leading to the formation of cycloalkene and CH$_3$ radical (reaction 4→10 in Schemes 3 and 4).

- The *endo* β-scission, that corresponds to the ring opening with the π-bond being formed within the ring (reactions 4→5, 4→11, 6→7 and 8→9 in Scheme 3 and reactions 3→7, 4→8, 4→9, 5→11 and 5→12 in Scheme 4). These reactions are similar to those considered in section 4, where no alkyl substitution was considered (§ 4).

- The *exo* β-scission, which corresponds to the ring opening with the π-bond being formed out of the ring (reaction 2 →3 in Scheme 3 and 2→6 in Scheme 4).

Consider the activation energies involved in the first series of reactions. We note that the values obtained are relatively close to the one reported for the β-scission of a CH$_3$ group in alkyl free radicals (31 kcal.mol$^{-1}$, Buda *et al.*[48]) and that the ring structure does not strongly affect the cleavage of the alkyl side chain. An interesting feature concerns the activation energy differences between *endo* and *exo* β-scissions. Indeed, *endo* β-scissions require activation energy greater than *exo* β–scissions (11 kcal.mol$^{-1}$ for methylcyclopentyl and 15 kcal.mol$^{-1}$ for methylcyclobutyl). **Figure 7** illustrates the TS geometry obtained at the B3LYP/cbsb7 level of theory for β-scissions in methylcyclobutyl radical: an *endo* β-scission (reaction 4→5 of Scheme 3) and an *exo* β-scission (reaction 2→3 of Scheme 3).

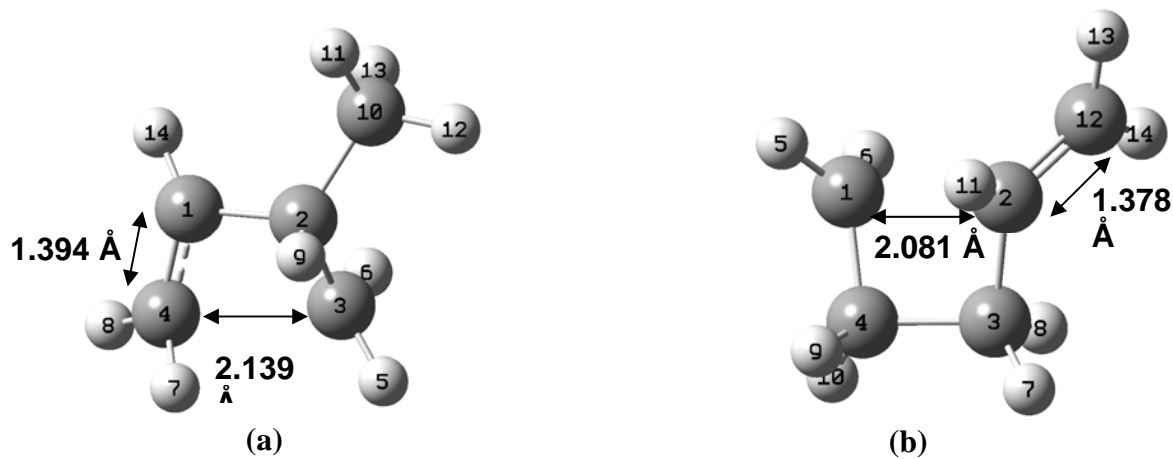

**Figure 7**: Geometry of the TSs in the β-scissions of methylcyclobutyl radicals, obtained at the B3LYP/cbsb7 of calculation. (a) *endo* ring opening (4 → 5 of scheme 3); (b) *exo* ring opening (2 → 3 of scheme 3).

As shown in Figure 7a, in the *endo* ring opening the formation of the π bond requires a rotation of the CH$_2$ group (atoms 4, 7 and 8) around the bond 1-4 in order to bring them in the same plane as atoms 1 and 14. As we have seen above, this rotation is hindered by the cyclic structure of the TS. On the contrary, in the *exo* β-scission (Figure 7b) formation of the π bond involves the rotation of the CH$_2$ group linked to the ring (atoms 12, 13 and 14) which can be carried out quite easily. Not surprisingly, *exo* β-scission exhibits lower activation energy than *endo* β-scission. C-C distances in Figure 7 also show that the distance between carbons 4 and 3 (d= 2.139 Å) in *endo* ring opening is longer than between carbons 1 and 2 (d= 2.081 Å) in *exo* ring opening. It is seen that in *endo* ring opening, the deformation of the cyclic structure is significant in order to create a double bond. It is worth noting that even in the *exo* β-scission, the ring strain energy is not totally released in the TS. In the case of the cyclobutyl radical, the activation energy would be 3.4 kcal.mol$^{-1}$ if all the RE had been released in the TS, whereas the value obtained for reaction 2→3 of Scheme 3 is equal to 12.3 kcal.mol$^{-1}$. This result confirms that the TS in the *exo* β-scission of methylcyclobutyl radical is tight with a pronounced cyclic character.

Another interesting comment can be made by comparing the activation energies and those obtained in the β-scission of straight-chain alkyl radicals. In the case of methylcyclobutyl radicals, the activation energies obtained are always lower, even for *endo* β-scissions, than those found for alkyl radicals (about 28 kcal.mol$^{-1}$). The release of a large part of the RE involved in cyclobutyl (RE = 24.4 kcal.mol$^{-1}$) balances with the steric inhibition created by the π bond formation. Conversely, for methylcyclopentyl radicals, RE is much smaller (4.1 kcal.mol$^{-1}$) and the partial recovery of RE is not sufficient to compensate the steric inhibition in the case of the *endo* β-scission. As a consequence, activation energies obtained for *endo* ring opening are greater that of straight-chain alkyl radicals.

### 5.2 Influence of the size of the lateral alkyl chain : ethylcyclopentyl decomposition

For ethylcyclopentyl, one observes the formation of the same type of products as in the methylcylopentyl radical, except for two new elementary processes presented on **Scheme 5**. They correspond to the rupture of the alkyl side chain and lead to the formation of cyclopentane-methylene and cyclopentyl radical. For reaction 1 → 2, the activation energy is close to that for the β-scission of the CH$_3$ group in a straight-chain alkyl radical[49], while the value obtained for the reaction 3 → 4 (23.9 kcal.mol$^{-1}$) is smaller. The less stable primary center in 3 is certainly responsible for the lowered activation energy as compared to the more stable, tertiary radical in 1.

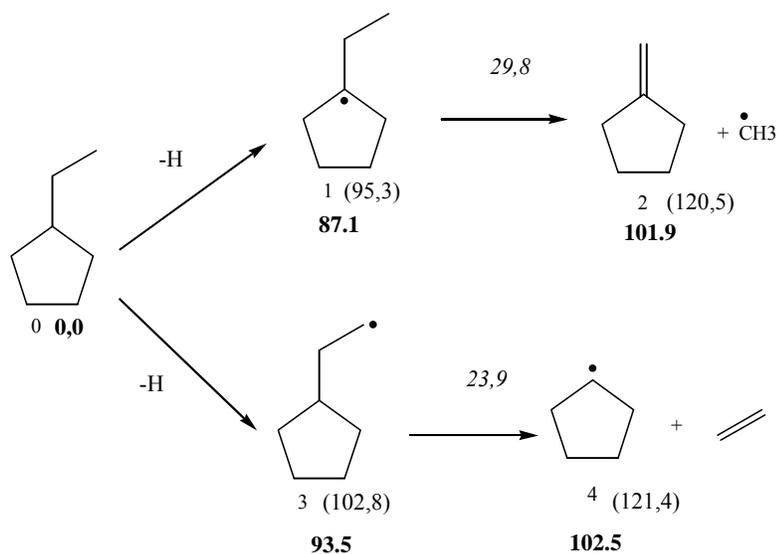

**Scheme 5**: β-scissions of two isomers of ethylcylopentyl radicals. Gibbs free energies (in bold), activation enthalpies (in italic) and enthalpies of reaction (in parentheses) are reported at 298 K, in kcal.mol$^{-1}$ and are relative to the reference cycloalkane.

As mentioned previously, all other reactions involve in the decomposition scheme of ethylclopentyl radicals lead to the same type of products with activation similar to those in methylcylopentyl radicals. Thus, the size of the alkyl side chain does not impact the activation energy for a given class of β-scission, i.e, *endo* or *exo* ring opening. The detailed scheme for ethylcylopentyl can be found in the supporting information.

### 5.3 Rate parameters

Kinetics parameters are presented in **Tables 8, 9** and **10**. It is interesting to note that for *exo* β-scission, the pre-exponential factor is lower than for *endo* β-scissions. This can be explained by an entropic effect since in the former case, the internal rotation of the CH$_3$ group is lost in the TS (due to the formation of the double bond), while in the latter case the rotation of CH$_3$ group contributes to the entropy of the TS.

**Table 8**: Rate parameters for the C-C bond breaking of methylcyclobutyl radicals in Scheme 3, $500 \leq T (K) \leq 2000$.

| Reaction | A (s$^{-1}$) | n | E (kcal.mol$^{-1}$) |
| --- | --- | --- | --- |
| 2 → 3 | 1.51×10$^{10}$ | 0.899 | 12.84 |
| 4 → 5 | 1.82×10$^{12}$ | 0.331 | 27.86 |
| 4 → 10 + CH$_3$ | 3.31×10$^{11}$ | 0.810 | 33.02 |
| 4 → 11 | 6.31×10$^{12}$ | 0.235 | 26.62 |
| 6 → 7 | 3.16×10$^{14}$ | -0.282 | 27.93 |
| 8 → 9 | 1.44×10$^{11}$ | 0.702 | 28.83 |

**Table 9**: Rate parameters for the C-C bond breaking of methylcyclopentyl radicals in Scheme 4, $500 \leq T (K) \leq 2000$.

| Reaction | A (s$^{-1}$) | n | E (kcal.mol$^{-1}$) |
| --- | --- | --- | --- |
| 2 → 6 | 2.75×10$^{9}$ | 0.991 | 23.26 |
| 6 → 2 | 7.93×10$^{4}$ | 1.951 | 6.45 |
| 3 → 7 | 1.86×10$^{12}$ | 0.419 | 34.37 |
| 4 → 8 | 3.47×10$^{10}$ | 0.959 | 33.91 |
| 4 → 9 | 4.27×10$^{12}$ | 0.404 | 33.71 |
| 4 → 10 + CH$_3$ | 4.57×10$^{11}$ | 0.843 | 30.53 |
| 5 → 11 | 1.44×10$^{13}$ | 0.277 | 33.74 |
| 5 → 12 | 6.76×10$^{11}$ | 0.692 | 35.45 |

**Table 10**: Rate parameters for the C-C bond breaking of ethylcyclopentyl radicals in Scheme 5, $500 \leq T$ (K) $\leq 2000$.

| Reaction | A (s$^{-1}$) | n | E (kcal.mol$^{-1}$) |
|---|---|---|---|
| 1→2 | 8.51×10$^9$ | 1.485 | 30.30 |
| 3→4 | 5.01×10$^8$ | 1.286 | 24.05 |

Very few rate constants are available in the literature for the β-scissions in methylcyclobutyl and methylcyclopentyl. Rate rules for cycloalkyl ring closures or openings were listed by Newcomb.[10] A rate constant for the *exo* ring opening of methycyclobutyl radical (reaction 2→3 of Scheme3) is proposed in an experimental study of Walton[50]. **Figure 8** compares the values obtained here for the rate constant with those proposed by Walton.

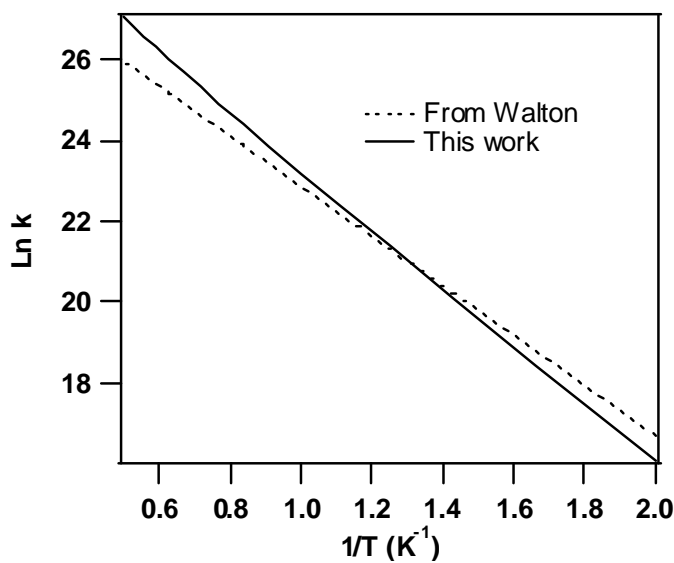

**Figure 8**: Comparison between the rate constant proposed by Walton[50] and that obtained from our CBS-QB3 calculations for the *exo* ring opening of methylcyclobutyl free radical (reaction 2→3 of Scheme 3), $500 \leq T(K) \leq 2000$.

It shows that the rate constant calculated in our work has a higher activation energy. However, the rate values are within a factor 0.5 to 3 of each other from 500 to 2000 K. For the *exo* β-scission of methylcyclopentyl, Newcomb[10] proposed a rate constant related to the inverse reaction, *i.e.* the ring closure of 1-hexen-6-yl radical, with rate parameters derived from room temperature experiments of Chatgilialoglu et al.[51] and estimation of pre-exponential factor based on analogous ring closure reactions. **Figure 9** compares the values obtained with the rate constant suggested by Newcomb and that obtained from our CBS-QB3 calculations for the reaction 6→ 2 related to Scheme 4.

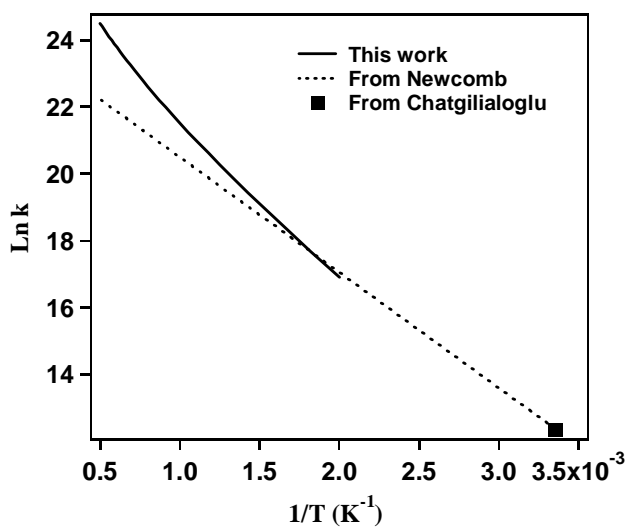

**Figure 9**: Comparison between the rate constant proposed by Newcomb and that obtained from our CBS-QB3 calculations (Table 9) for the ring closure of 1-hexen-6-yl radical (reaction 6→ 2 of Scheme 4). $500 \leq T(K) \leq 2000$.

As seen, the rate constants are in good agreement at around 800 K. At 2000 K, a factor 10 difference is reached. Since the rate expression of Newcomb was derived from room-temperature experiments of Chatgilialoglu et al.[51], as shown in Figure 9, the discrepancy at high temperature is not surprising.

## 6. Evans-Polanyi correlation for the β-scission of cycloalkyl and linear alkyl free radicals

Our theoretical study on the ring opening of cycloalkyl radicals highlights the roles of the ring strain energy and the π-bond formation in the transition state. We showed that three types of reactions can occur during the decomposition of cycloalkyl radicals: *endo* and *exo* ring opening and β-scission of an alkyl group. It is interesting to compare activation energies involved in the three elementary β-scission processes with the enthalpies of reaction, i.e., to build an Evans-Polanyi plot at 298 K. The plot has been performed by considering all the reactions previously discussed in Scheme 2, 3 and 4. We also included further calculations of the activation energies for all possible ring opening processes (*endo* and *exo*) in methylcyclopropyl and ethylcyclopentyl isomers, as well as β-scissions of the alkyl side chain. All of these results have been obtained at a CBS-QB3 level of calculation.

In order to compare the activation energies involved in the β-scissions of cycloalkyls to those of linear alkyl free radicals in the Evans-Polanyi plot, we have also considered the rupture of C-C and C-H bonds for several linear alkyl radicals presented in **Table 11**.

**Figure 10** shows the results obtained for the Evans-Polanyi plot at 298K and for all the reactions mentioned above. Activation energies for β-scissions of alkyl groups, C-H ruptures and *exo* ring openings of cycloalkyl radicals correlated well with C-C and C-H bond breaking of linear alkyl radicals, except for β-scissions #1, #2 and #3 (Figure 10). These last reactions correspond to *exo* β-scissions of methyl cyclopropyl radicals (#1, #2) and C-H bond scission of cyclopropyl (#3).

These results are not surprising since the RE in cyclopropyl and cyclopropene are very important and can lead to particular behaviour. The activation energy obtained for *endo* ring opening in the $C_6$ cycloalkyl radical is also in good agreement with the previous ones. The latter result can be explained by the low ring strain energy involved in this structure, associated with a moderate steric inhibition effect when the π bond is being formed in the TS. For $C_7$ ring opening the agreement is a little worse and the activation energy obtained has not been included in the correlation.

**Table 11**: Reaction enthalpies and activation enthalpies obtained at the CBS-QB3 level of calculation, for the β-scissions of C-C and C-H bonds of acyclic radicals and used in the Evans-Polanyi plot (T=298 K).

| Reactions considered in Evans-Polanyi correlation | $\Delta_r H°$ (kcal/mol) | $\Delta H^{°\neq}_{298K}$ (kcal/mol) |
|---|---|---|
| $•C_3H_7 \rightarrow •CH_3 + C_2H_4$ | 22.8 | 28.9 |
| $•C_4H_9 \rightarrow •C_2H_5 + C_2H_4$ | 22.0 | 27.8 |
| $•C_5H_{11} \rightarrow •C_3H_7 + C_2H_4$ | 22.8 | 28.2 |
| $•C_6H_{13} \rightarrow •C_4H_9 + C_2H_4$ | 22.7 | 28.0 |
| $•C_7H_{15} \rightarrow •C_5H_{11} + C_2H_4$ | 22.7 | 28.0 |
| 1-hexen-6-yl $\rightarrow$ 1-buten-4-yl + $C_2H_4$ | 22.8 | 27.8 |
| 1-hexen-3-yl $\rightarrow$ buta-1.3-diene + $•C_2H_5$ | 15.7 | 25.0 |
| 1-penten-5-yl $\rightarrow$ $•C_3H_5 + C_2H_4$ | 7.4 | 19.2 |
| $•C_3H_7 \rightarrow C_3H_6 + H$ | 32.0 | 34.0 |
| $•C_4H_9 \rightarrow C_4H_8 + H$ | 32.6 | 34.4 |
| $•C_5H_{11} \rightarrow C_5H_{10} + H$ | 32.5 | 34.4 |
| $•C_6H_{13} \rightarrow C_6H_{12} + H$ | 32.4 | 34.4 |
| $•C_7H_{15} \rightarrow C_7H_{14} + H$ | 32.4 | 34.2 |

**Figure 10**: Evans-Polanyi correlation for the β-scissions of linear and cyclic alkyl free radicals at T= 298 K. Values obtained at the CBS-QB3 level of calculation.

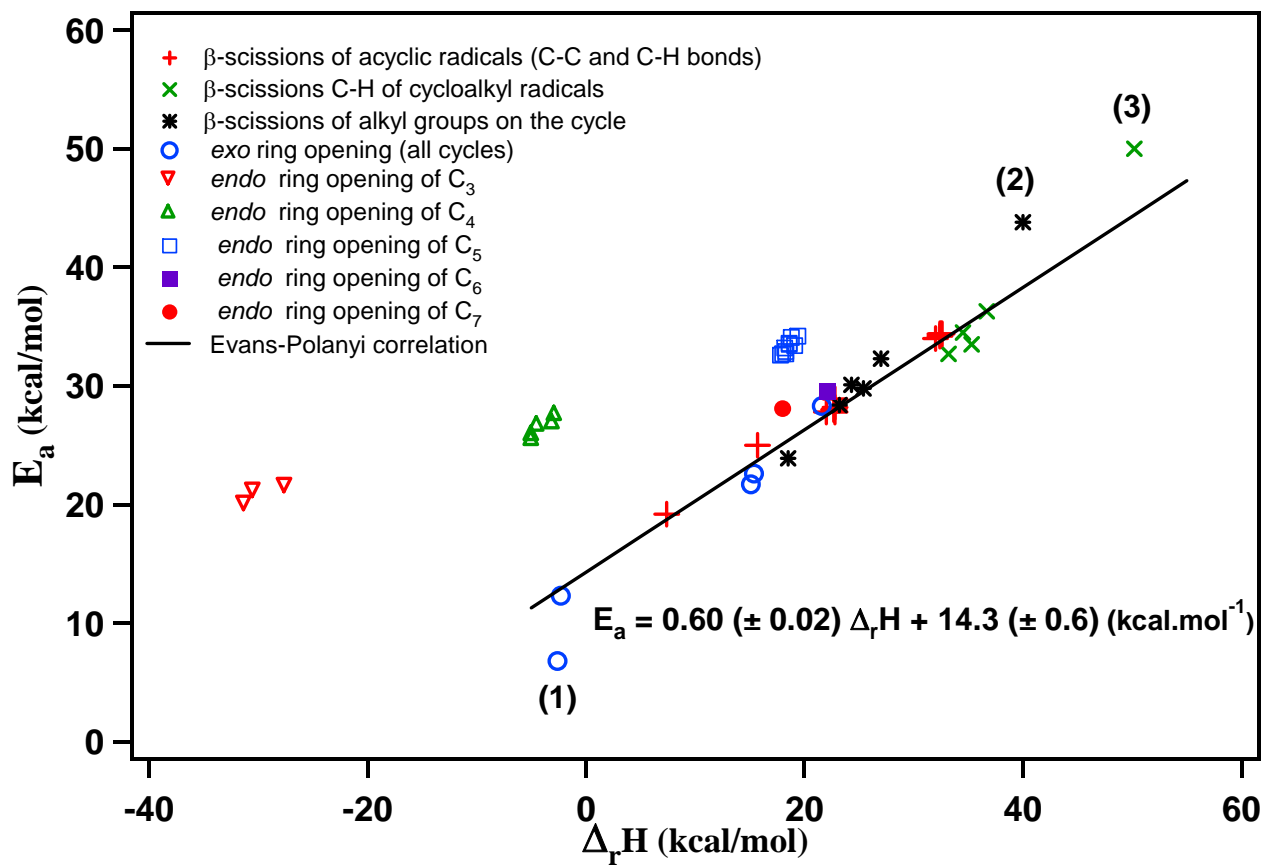

For *endo* ring opening of $C_3$, $C_4$ and $C_5$ cycloalkyl radicals, activation energies are clearly not correlated with β-scissions of straight-chain alkyl radicals. However, we can note that for a given size of the cyclic structure, the effect of alkyl substitution is relatively weak and, at a first approximation, the same activation energy can be used in this case. The Evans-Polanyi correlation (equation 8) was obtained by including all reactions considered in this study except those of *endo* ring opening of $C_4$, $C_5$ and $C_7$ cyloalkyl radicals and all C-C and C-H bond cleavage of $C_3$ cyclic structures:

$$E_a \text{ (kcal.mol}^{-1}\text{)} = 14.3 \ (\pm 0.6) + 0.60 \ (\pm 0.02) \ \Delta_r H°_{298K} \text{ (kcal.mol}^{-1}\text{)} \quad (8)$$

The latter simplification has been done in order to obtain a more accurate Evans-Polanyi correlation but also because of practical importance of the $C_4$, $C_5$ and $C_6$ cyclic structures.

For *endo* ring opening of $C_4$ and $C_5$ cycloalkyl free radicals, the rate parameters given in Table 6 can be used as estimates in mechanism construction.

7. Conclusion

In this paper, unimolecular decomposition of cycloalkyl radicals with/without an alkyl side chain were investigated at the CBS-QB3 level of theory. The study of unsubstituted cycloalkyl radicals revealed the importance of two opposite effects on the activation energy ($E_a$). The first one tends to decrease the energy barrier and is related to the ring strain energy (RE) in the reactants. In general, only a part of the RE is released at the saddle point. The second effect tends to increase $E_a$ and is related to π bond formation in the TS, which introduce steric hindrance. This effect is particularly important in the case of cyclopentyl ring. $E_a$ for this reaction is as high as 33.5 kcal.mol$^{-1}$. Different methylcycloalkyl radicals were also investigated. The radical center is located either in the ring carbon atom or on the alkyl side chain. If the radical center is in the ring, two types of reaction can occur: a) an *endo* ring opening, which presents the same characteristics as the equivalent process in unsubstituted cycloalkyl radicals, and b) the removal of the alkyl side chain. If the radical center is on the methyl side chain, *exo* ring openings can occur that proceed with a lower $E_a$ than *endo* ring openings. This can be explained by the fact that in *exo* ring opening, the formation of the π bond occurs out of ring, involving a smaller steric inhibition. Increasing the size of the alkyl side chain from methyl to ethyl leads to additional pathways involving the fragmentation of the alkyl chain. The ring opening in branched cycloalkyl radicals appears to be independent of the size of the alkyl side chain. The C-H bond breakings for cycloalkyl radicals have been studied and it has been shown that the C-H β-scission rate is close to that of a strain-free alkyl.

Finally, an Evans-Polanyi correlation is proposed for β-scissions of straight-chain alkyl radicals, *exo* ring opening in $C_4$ to $C_7$, C-H and alkyl chain ruptures of $C_4$, $C_5$ and $C_6$ cycloalkyl radicals, and *endo* $C_6$ ring opening. For *endo* ring opening in $C_3$, $C_4$ and $C_5$ cycloalkyl radicals, the Evans-Polanyi diagram shows that a lateral alkyl side chain does not affect the activation energy involved in β-scissions strongly. The Evans-Polanyi correlation derived in this work constitutes an interesting contribution in the field of cyclic alkane reactivity as it allows the estimation of the activation energy of C-C and C-H bond β-scission involving not only linear alkyl radicals but also any cyclic alkyl radicals with a ring size ranging from $C_4$ to $C_6$, with or without alkyl substitution.

ACKNOWLEDGMENT

We thank Professor A. Hocquet (Laboratoire de Chimie Physique Macromoléculaire of Nancy Universités. France) for helpful discussions concerning AIM analyses. The Centre Informatique National de l'Enseignement Supérieur (CINES) is gratefully acknowledged for allocation of computational resources.

SUPPORTING INFORMATION AVAILABLE

The full list of author in ref 16., the structural parameters for all the species investigated in this study, energies and zero point energies, isodesmic reactions and vibrational frequencies are provided in the supporting information. The detailed decomposition schemes of methylcyclopropyl and ethylcylopentyl are also given. This material is available free of charge via the Internet at http://pubs.acs.org.